# Moment-Tensor-Based Constant-Potential Modeling of Electrical Double Layers

Zhenxiang Wang[1,4], Ming Chen[1,4], Jiedu Wu[2], Xiangyu Ji[1], Liang Zeng[1], Jiaxing Peng[1], Jiawei Yan[2], Alexei A. Kornyshev[3], Bingwei Mao[2], and Guang Feng[1*]

[1]State Key Laboratory of Coal Combustion, School of Energy and Power Engineering, Huazhong University of Science and Technology, Wuhan 430074, China
[2]State Key Laboratory of Physical Chemistry of Solid Surfaces, and Department of Chemistry, College of Chemistry and Chemical Engineering, Xiamen University, Xiamen, China.
[3]Department of Chemistry, Faculty of Natural Sciences, Imperial College London, Molecular Sciences Research Hub, White City Campus, W12 0BZ, London, UK.
[4]These authors contributed equally: Zhenxiang Wang, Ming Chen.
[*]Correspondence: gfeng@hust.edu.cn

Constant-potential molecular dynamics (MD) simulations are indispensable for understanding the capacitance, structure, and dynamics of electrical double layers (EDLs) at the atomistic level. However, the classical constant-potential method, relying on the so-called 'floating charges' to keep electrode equipotential, overlooks quantum effects on the electrode and always underestimates EDL capacitance for typical electrochemical systems featuring metal electrodes in aqueous electrolytes. Here, we propose a universal theoretical framework as *moment-tensor-based constant potential method* (mCPM) to capture electronic structure variations with electric moments. For EDLs at Au(111) electrodes, mCPM-based MD reveals bell-shaped capacitance curves in magnitude and shape both *quantitatively* consistent with experiments. It further unveils the potential-dependent local electric fields, agreeing with experimental observations of redshift vibration of interfacial water under negative polarization and predicting a blueshift under positive polarization, and identifies geometry dependence of two time scales during EDL formation.

***Introduction.*** – The electrical double layer (EDL) at the electrode-electrolyte interface is ubiquitous in electrochemistry[1]. Delving into the intricate microstructure and dynamics of EDLs at the molecular scale helps to reveal pivotal mechanisms that determine electrochemical device performance, paving the way for transformative breakthroughs in advanced applications including electric energy storage, electrocatalysis, and capacitive deionization[2-4]. Molecular dynamics (MD) simulations, with their ability to sample phase space[5], have emerged as an indispensable tool for scrutinizing such nanoscale interfaces. A salient challenge in employing MD simulations for EDLs lies in adequately capturing the electronic response of electrodes to external fields[6,7], relying on an extended degree of freedom on the electronic structure. The constant potential method (CPM), with fluctuating charges on nuclei of electrode atoms subject to an equipotential constraint on the electrode, effectively reflects varying electrode electronic structures[8,9] and has been extensively utilized in elucidating both equilibrium and dynamic processes in various electrochemical systems[10,11].

A long-standing issue for MD simulations with classical CPM (cCPM) is the severe underestimation of EDL capacitance in typical electrochemical systems with metal electrodes[1,12,13]. The sharp discrepancy suggests the incomplete description of electrified solid-liquid interfaces due to classical electrostatics for cCPM, omitting quantum effects on electrodes[14,15]. Modifications based on a semi-classical model, linking the self-energy of electrode atoms to metallicity, show that decreasing the electrode atom 'hardness' could increase the capacitance; however, an overly low hardness would cause unstable simulation due to the polarization catastrophe[16]. *Ab initio* molecular dynamics (AIMD) modeling with the rigorous description of the electronic structure indicates that high EDL capacitance depends strongly on the interfacial dipole induced by chemisorbed water[17], which cannot be observed by classical MD simulations[17]. Nevertheless, *ab initio* methods are still not well applicable to either an overall EDL structure or charging dynamics because of the limited spatial and temporal scales[15].

In this Letter, a theory for cCPM is promoted by introducing multipole moment tensors to describe the variation in the electrode electronic structure (*i.e.,* induced charges[18]), termed moment-tensor-based CPM (mCPM). The moment tensors in mCPM are extracted from the spatial distribution of induced charges through density functional theory (DFT) calculations. We then conducted comparative analyses involving the cCPM and developed mCPM on the capacitance, structure, and dynamics of EDLs at interfaces between Au(111) electrodes and aqueous electrolytes. The mCPM-based MD predicts a bell-shaped potential-capacitance curve with a much higher magnitude than cCPM, aligning quantitatively with experimental measurements. It further unveils the origin of the water and ions in EDLs, responding to the external field, and observes the two-stage charging dynamics process composed of ion electromigration and bulk diffusion.

*From cCPM to mCPM.*– The distribution of induced charges on the electrode inherently exhibits non-uniformity [Fig. 1(a)], dynamically responding to perturbations from electrolyte ions and solvent molecules[6]. However, cCPM, adopting nucleus-centered charges, only describes the magnitude of induced charges and neglects the electrostatic interactions due to asymmetric electronic structure[19]. To give a complete description of electrostatic interactions involving induced charges, both net charge and electric moments are required according to the multipole expansion of the electrostatic potential. Therefore, mCPM with moment tensors is proposed to depict electrostatic interactions from *both* the amount and the distribution of induced charges. Specifically, induced charges are divided and expanded into $N$ multipoles, each consisting of fluctuating multipole moment tensors [Fig. 1(b)] as

$$\boldsymbol{M}^{(i)} = \{Q^{(i)}, \boldsymbol{\mu}^{(i)}, \boldsymbol{\Theta}^{(i)}, \ldots\} \tag{1}$$

where $Q^{(i)}$ is the $i^{th}$ monopole (net charge), and $\boldsymbol{\mu}^{(i)}$, $\boldsymbol{\Theta}^{(i)}$ are the dipole, and quadrupole moment tensors, respectively.

As only the $0^{th}$ rank tensors $\boldsymbol{Q} = \{Q^{(1)}, \ldots, Q^{(N)}\}$ were employed to describe the electrostatic effect of the net charges, neglecting the effects of electric moments, Eq. 1 could reduce to the theory of cCPM, so that cCPM is a 'first approximation' to the problem. In comparison, the induced charges in mCPM are represented by multipole tensors $\boldsymbol{M} = \{\boldsymbol{M}^{(1)}, \ldots, \boldsymbol{M}^{(N)}\}$, including both net charges and electric moments. Therefore, within this framework, the extended Hamiltonian of the MD simulation system becomes[14]:

$$H(\boldsymbol{r}, \boldsymbol{p}, \boldsymbol{M}) = T(\boldsymbol{p}) + U_{coul}(\boldsymbol{r}, \boldsymbol{M}) + U_0(\boldsymbol{r}) \tag{2}$$

where $T$ is the kinetic energy, $U_{coul}$ is the electrostatic energy, $U_0$ includes bond energy and van der Waals energy; $\boldsymbol{r}$ and $\boldsymbol{p}$ are the positions and momenta of the atoms, respectively. The moment tensors are solved by minimizing $U_{coul}(\boldsymbol{r}, \boldsymbol{M}) - \sum_{i=1}^{N} \Psi^{(i)} Q^{(i)}$ to satisfy the equipotential constraints, $\Psi^{(i)}$, on the electrode.

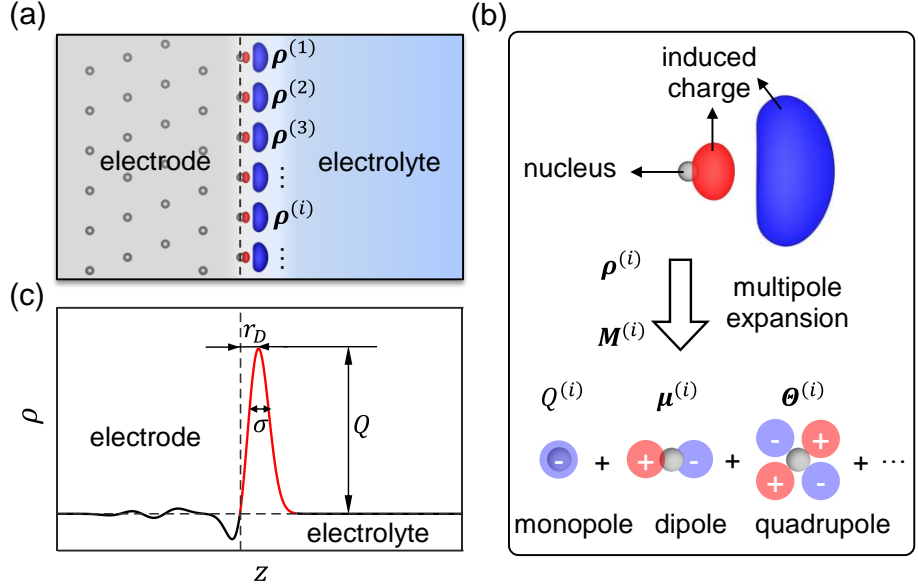

FIG. 1. Schematics of mCPM. (a) Schematic of induced charge in electrode-electrolyte interface. Gray spheres represent electrode nuclei. (b) Multipole expansion of induced charge. Red and blue isosurfaces refer to the positive and negative induced charges, respectively. (c) Schematic of induced charge distribution in metal electrodes and electrolytes.

The mCPM theory constitutes a versatile framework to characterize the induced charges efficiently and completely, which is applicable to diverse electrochemical systems. As this Letter focuses on metal-liquid interfaces, the combined effect of monopole and dipole could sufficiently describe the electrostatic effect of induced charges[18,20]. Therefore, each moment series is denoted as $\boldsymbol{M}^{(i)} = \{Q^{(i)}, \boldsymbol{\mu}^{(i)}\}$, and the electrostatic potential at a specific location **r** due to the electrical moments can be written as:

$$\phi(\mathbf{r}) = \sum_{i=1}^{N} \left( \frac{Q^{(i)}}{r} - \frac{\boldsymbol{\mu}^{(i)} \cdot \mathbf{r}}{r^3} \right) \quad (3)$$

where

$$\boldsymbol{\mu}^{(i)} = Q^{(i)} \mathbf{r}_D \quad (4)$$

$Q^{(i)}$ and $\boldsymbol{\mu}^{(i)}$ are the monopole and dipole of the $i^{\text{th}}$ part of the induced charge, respectively, and $\mathbf{r}_D$ is the displacement vector of the dipole. Herein, the monopole term is represented by Gaussian charge, $\rho^{(i)}(\mathbf{r}) = Q^{(i)}(\pi\sigma^2)^{-3/2} e^{-\mathbf{r}^2/\sigma^2}$, with a width, $\sigma$, reflecting the atom hardness [Fig. 1(c)]. Detailed theoretical derivation is given in Section 1 of the Supplemental Material (SM).

To obtain $\sigma$ and $\mathbf{r}_D$ for the electrochemical system with Au(111) electrode, DFT calculations are performed based on both explicit ion absorption and implicit solution (Section 2 of SM). The induced charges on the metal electrode exhibit a similar shape under different polarization conditions, providing the distribution width (0.042 nm) and displacement vector of the dipole (0.102 nm pointing out straight from electrode surface) for the Au(111) electrode (details see Section 3 of SM and Figs. S2-3).

***EDL capacitance and structure.*** – The EDL capacitance was examined in a benchmark electrochemical system consisting of 2 M $NaClO_4$ electrolyte confined between two atomic flat Au(111) electrodes (Fig. S4). As illustrated in Fig. 2(a), cCPM-MD yielded an almost flat differential capacitance profile at approximately 6 µF $cm^{-2}$, consistent with prior CPM-MD simulations[12]. In contrast, a distinct bell-shaped curve with a significantly higher magnitude was found with mCPM [Fig. 2(a)]. Meanwhile, our electrochemical impedance spectroscopy measurements on single-crystal Au(111) electrodes with the same electrolyte (Section 2 of SM) also demonstrate a bell-shaped curve ranging from 24 to 41 µF $cm^{-2}$, quantitatively consistent with mCPM predictions and compatible with previous experiments on Au and Pt electrodes[21]. These agreements prove the pivotal role of electric moments in reshaping the differential capacitance.

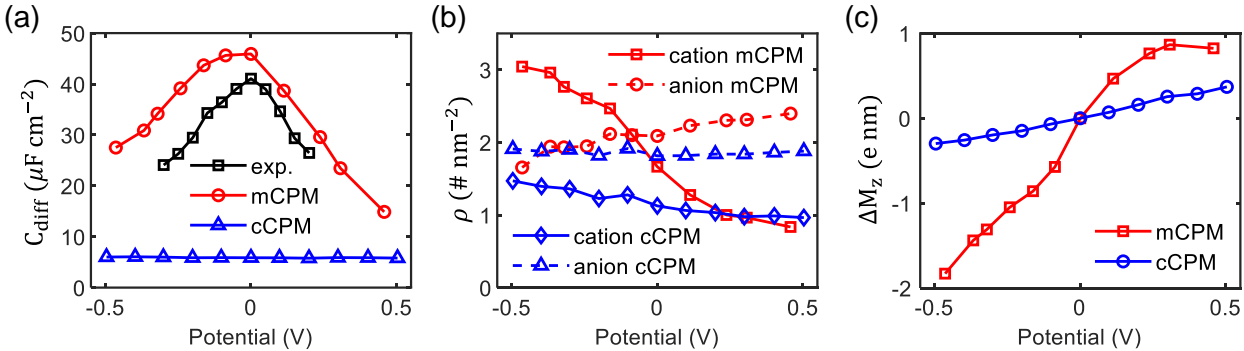

FIG. 2. Differential capacitance and its origin. (a) Potential dependence of the capacitance of Au(111) in 2 M $NaClO_4$ solution derived from experiments, mCPM, and cCPM. (b) Accumulative number densities of the first ion layer under various potentials. (c) Excess polarization of interfacial water at different potentials. 0 V refers to the potential of zero charge.

The origins of the differential capacitance curve could be ascribed to the EDL structure, as earlier theoretical studies ascribe the capacitance hump around the potential of zero charge to the reorientation of interfacial water molecules[22], while the mean field theory suggested that the close-packed ions and solvents would result in nonlinear ion absorption and therefore nonmonotonic differential capacitance[23,24]. Therefore, the potential-dependent ion absorption/desorption and interfacial water polarization are compared thoroughly between cCPM and mCPM to understand their differential capacitance curves. In cCPM, the accumulative number of cations in the EDL exhibits a nearly linear variation under different potentials, with anions remaining almost unchanged [Fig. 2(b) and Fig. S5(a,b)]. Simultaneously, the interfacial water, acting as dielectric medium to screen electric field and contribute to charge storage[25], exhibits linear growth of excess polarization with increasing potential [Fig. 2(c)]. These weaker and linear responses lead to a flat and small capacitance. In sharp contrast, mCPM reveals a fundamentally different charge storage mechanism that cation absorption and desorption are much stronger than those in cCPM [Fig. 2(b) and Fig. S5(c,d)]. The interfacial water also shows stronger excess polarization under negative polarization and weakly positive polarization (<0.3 V) compared to cCPM [Fig. 2(c)]. These distinct EDL structures account for the much higher capacitance observed in mCPM. Besides, the variation in interfacial cation number weakens under stronger polarization [Fig. 2(b)] and excess polarization of interfacial water grows subtly when the potential exceeds 0.3 V [Fig. 2(c)], decreasing the capacitance. Therefore, induced electric moments are indispensable in determining the EDL structures, and the synergistic effects of ion absorption/desorption and water polarization contribute to the more precise bell-shaped capacitance.

Notably, previous MD simulations[26,27] also reproduced non-linear capacitance curves with the modified constant charge method (CCM) where the electrode polarization is represented by off-center constant charges. The change in capacitance only arises from the different reference plane of the electrode potential (Fig. S6-7), rather than different EDL structures, proving that off-center CCM cannot reflect the variations in the electronic structures of electrodes.

***Interfacial water.***– Potential-dependent water structure at the charged surface is crucial for understanding interfacial phenomena[28,29]. In cCPM-MD, water absorption remains unaffected by the electrode potential [Fig. 3(a)]. In comparison, the water layer observed in mCPM-MD is strongly dependent on the electrode potential. A notable amplification in interfacial water adsorption is observed with increasing negative polarization, displaying closer adsorption and nearly tripling the height of the first adsorbed peak from 0 V to -0.5 V [Fig. 3(b) and Fig. S8(a-e)]. This pronounced effect can be attributed to the strong absorption of hydrated cations (Fig. S5).

Regarding orientational distribution, cCPM elucidates a water configuration nearly parallel to the electrode surface with negligible dependence on electrode polarization[Fig. 3(c), and Fig. S8-9]. In contrast, mCPM reveals a markedly distinct motif for interfacial water [Fig. 3(d)]. Under 0 V, the dipole orientation of water molecules displays two peaks (at ~105° and ~130°), indicating a coexisting parallel and perpendicular orientation with one hydrogen atom pointing to the surface (*i.e.*, H-down configuration). As the potential becomes negative, more water molecules exhibit an H-down structure. Specifically, interfacial water with parallel configuration almost disappears at potentials below -0.3 V [Fig. 3(d)], with only a narrow peak of dipole orientation located at ~130°, closely aligning with AIMD simulation[17] and SHINERS experiment[29]. These findings prove that both strong water electrosorption and water polarization exist under negative polarization in mCPM, which is prone to water electrolysis[30], while cCPM gives a contrary hint. As the applied potential turns positive, the dipole orientation of water molecules undergoes a transition from ~130° to 109° and eventually drops below 90° [Fig. 3(d)], indicating the occurrence of O-down configuration, consistent with previous sum frequency generation measurements and DFT calculations[31,32]. Detailed interfacial water structures see Section 6 of SM.

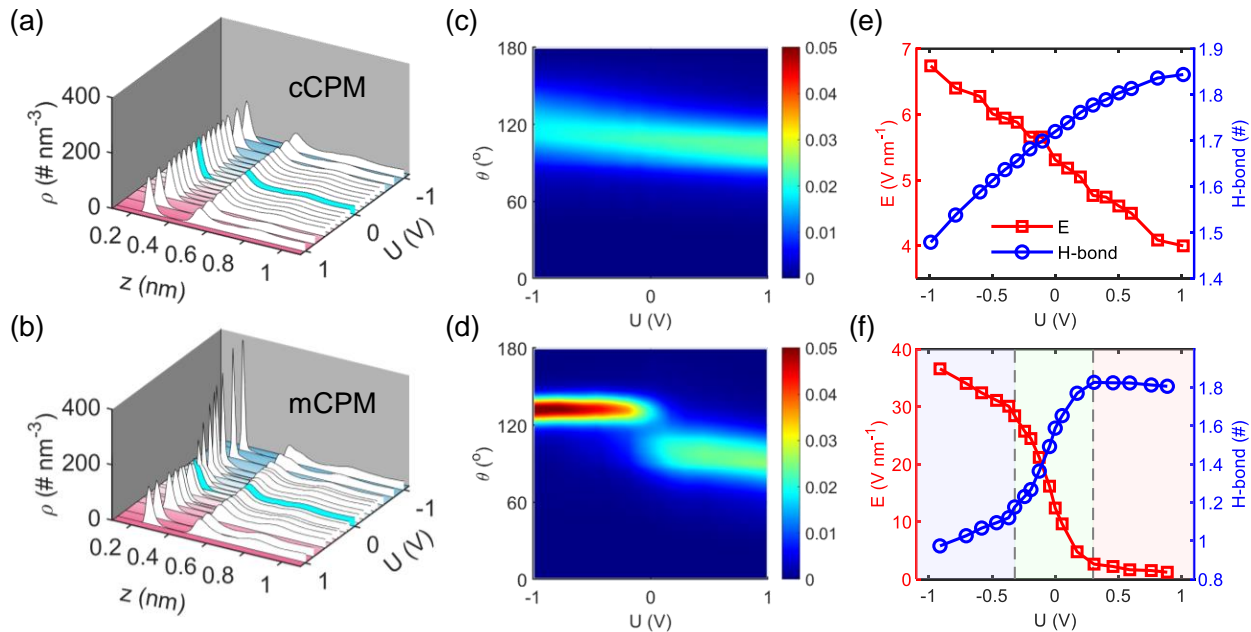

FIG. 3. Structure of interfacial water and electric field. (a-b) Number densities of water as a function of distance from the electrode surface nuclei (z) under various potentials (U). (c-d) Dipole orientations of interfacial water. $\theta_{dipole}$ is defined as the angle between the normal of electrode surface and the water vector. (e-f) Interfacial electric field and H-bond of interfacial water. Top and bottom panels refer to cCPM and mCPM, respectively.

It is of particular interest to uncover the potential-dependent local electric field experienced by interfacial water, which correlates with OH-stretching frequency shifts based on Stark effect[33]. cCPM exhibits a nearly linear increase in the magnitude of local electric field, and a linear growth in the number of H-bonds with growing electrode potential [Fig. 3(e)], corresponding with the linear variation in excess polarization and cation absorption/desorption. Intriguingly, mCPM reveals two transitions in the slope of the local electric field at approximately ±0.3 V, delineating three Stark tuning ranges [Region I: -0.3 ~ 0.3 V; Region II: < -0.3 V; Region III: > 0.3 V, as delineated in Fig. 3(f)]. The transition at -0.3 V is in accordance with the SHINERS experiment on Pd/Au-NaClO$_4$ interfaces, identifying two Stark tuning rates for the interfacial water OH stretch mode with the transition at -0.31 V[29]. Furthermore, the slope of the potential-dependent local electric field in Region I is steeper than in Region II, consistent with SHINERS measurements indicating a greater redshift in Region I than in Region II[29]. Simultaneously, the number of H-bonds exhibits a similar potential-dependent trend, quickly decreasing in Region I and subsequently gradually decreasing below -0.3 V (Region II). These transitions coincide with the vanishment of water molecules with parallel configuration under strongly negative polarization, which is essential to understanding the electrode reaction path and rate[34]. Under positive polarization, the local electric field decreases with increased polarization, implying a blueshift in OH stretching mode. The transition in the slop at 0.3 V signifies a Stark tuning rate transition, necessitating experimental verification.

*Charging dynamics.*– The dynamics of EDL formation determine the power density of electrochemical devices[35]. Typically when a voltage is applied, there are three time scales of the charging process[36]. Earlier theoretical studies propose that the charging process consists of the Debye time ($\lambda_D^2/D$) associated wtih the relaxation within EDL and the *RC* time ($\lambda_D L/D$) representing ions entering/leaving the EDL[37], where $\lambda_D$ is Debye length, $D$ is ionic diffusivity, and $L$ is half of the electrode separation. Recent analytical solutions to the Poisson-Nernst-Planck (PNP) equations introduce the bulk diffusion time ($L^2/D$) reflecting bulk electrolyte filling in the depleted zones after ion electromigration in the *RC* time, and conclude that the Debye time is a small perturbation in the *RC* time which dominates the charging process of thin EDLs ($L \gg \lambda_D$)[36,38]. Notably, the charging process revealed by molecular simulation are several orders of magnitude faster than experiments[39], making it urgent to accurately capture relevant time scales and their geometry dependence.

We scrutinized the charging dynamics of the EDL system of two parallel Au(111) electrodes embedded into 2 M NaClO$_4$ electrolyte ($\lambda_D \approx 0.2$ nm and $D \approx 1.2\times10^{-9}$ m$^2$ s$^{-1}$, Fig. S4) with electrode separations of 10, 30, 60, and 100 nm under a voltage jump of 1 V by both cCPM and mCPM. Simulated by cCPM [Fig. 4(a)], an immediate charge accumulation occurs within ~10 ps after the voltage jump, succeeded by a more gradual charging process. The latter becomes dominant with larger electrode separation exceeding 30 nm [Fig. S10(a)]. Conversely, by mCPM [Fig. 4(b) and Fig. S10(b)], the charging curves are composed of a typical double exponential process with a longer relaxation time, where the initial stage constitutes the predominant component of the net charge.

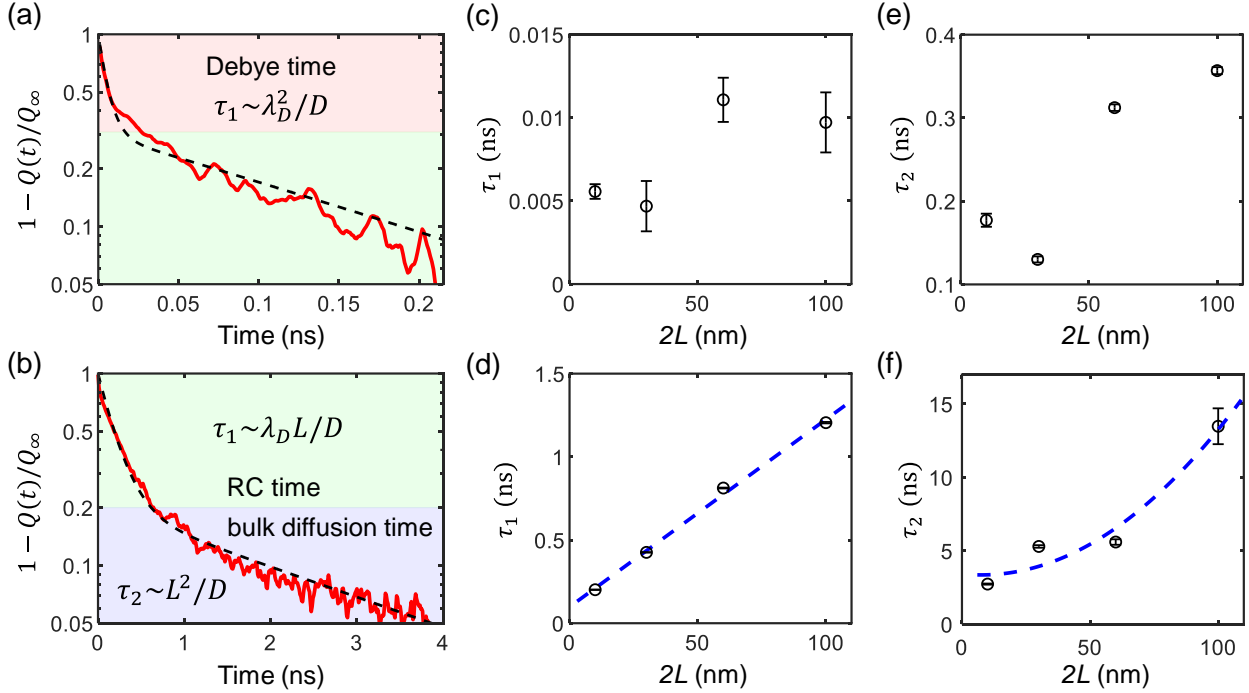

FIG. 4. Charging process. (a-b) Time evolution of net charge at positive electrode after a voltage jump of 1V applied for systems with 10 nm electrode separation. Black dashed lines represent double exponential fitting results. (c-d) Fast time scales with different electrode separations. Blue dotted line is a linear fitting result ($R^2 = 0.998$). (e-f) Slow time scales with different electrode separations. Blue dotted curve is a parabolic fitting result ($R^2 = 0.951$). Top and bottom panels refer to cCPM and mCPM, respectively.

To quantitatively investigate the two-stage charging, the charging curves are fitted with the double exponential function[36]:

$$Q(t) = Q_\infty \left[1 - A\exp\left(-\frac{t}{\tau_1}\right) - (1-A)\exp\left(-\frac{t}{\tau_2}\right)\right] \quad (5)$$

where $Q_\infty$ is the charge density at equilibrium, $\tau_1$ and $\tau_2$ are time scales ($\tau_1 < \tau_2$), and $A$ is the weight coefficient of the fast time scale. For cCPM-MD [Fig. 4(c,e)], $\tau_1$ approaches the Debye time and changes subtly with electrode separation, suggesting dominance by rearrangement within EDL. $\tau_2$ is at the level of $RC$ time but it does not grow linearly with $L$, incompatible with any theoretical predicted time scales. The fast stage still accounts for more than 20% of the total charge storage with wide separations (Table S1), probably due to the underestimated capacitance by cCPM that amplifies perturbations from EDL rearrangement. For mCPM-MD [Fig. 4(d,f)], $\tau_1$ increases linearly with $L$ [Fig. 4(d)], indicating dominance by ion electromigration. Meanwhile, $\tau_2$ grows proportionally to the square of the electrode separation, suggesting it is in the slow bulk diffusion process. These are typical characteristics of thin EDLs according to the PNP model[36,38], proving the accuracy of mCPM-MD in EDL charging dynamics. Moreover, the weight coefficients demonstrate that the charging process in these systems cannot be modeled by an *RC* circuit regardless of the thin EDL condition, as the bulk diffusion process constitutes ~15% of total charge storage and remains stable with increasing separation (Table S1). Considering both *RC* time scale and bulk diffusion time scale, predictions of the power performance of electrochemical devices become achievable through mCPM-MD.

***Discussion.*** – In this Letter, the developed mCPM-MD theory and method, utilizing multipole moment tensors to express electrostatic interactions from DFT-derived induced charges on electrodes, could effectively address the long-standing obstacle of quantitative agreement between experiments and atomistic modeling of EDLs[12-15]. As for a prototype electrochemical system of Au(111) electrodes in aqueous electrolytes, mCPM-MD results quantitatively match experimental bell-shaped capacitance curves and exhibit potential-dependent local electric fields consistent with experimental redshift vibration of interfacial water[29]. Meanwhile, mCPM-MD notably predicts a blueshift in the water vibration under positive polarization, and also identifies the geometry-dependent time scales in the charging process between parallel-plate electrodes, corresponding to ion electromigration and bulk diffusion. This bridges the time scale between molecular simulations and macroscopic devices, providing valuable insights into EDL evolution.

This Letter advances the CPM theory to effectually consider quantum effects on the electrode, and then enables modeling EDL system in millions of atoms and capturing variations in electronic structures, taking advantage of classical MD simulation and DFT calculations. It also aids in understanding interfacial phenomena of other EDL-related fields, such as batteries[4], electrocatalysis[2], and capacitive deionization[3].

The authors in China acknowledge the funding support from the National Natural Science Foundation of China (T2325012, 52106090, 52161135104, and 22072123) and the Program for HUST Academic Frontier Youth Team. A.A.K. thanks the grant from the Engineering and Physical

Sciences Research Council (EP/L015579/1).[1] W. Schmickler, Chemical Reviews **96**, 3177 (1996).

[2] M. F. Döpke, F. W. v. d. Meij, B. Coasne, and R. Hartkamp, Physical Review Letters **128**, 056001 (2022).

[3] R. A. Rica, R. Ziano, D. Salerno, F. Mantegazza, and D. Brogioli, Physical Review Letters **109**, 156103 (2012).

[4] J.-Q. Huang, Nature Nanotechnology **17**, 680 (2022).

[5] M. Salanne, J. M. Buriak, X. Chen, W. Chueh, M. C. Hersam, and R. E. Schaak, ACS Nano **17**, 6147 (2023).

[6] A. Kornyshev, Electrochimica Acta **34**, 1829 (1989).

[7] G. Jeanmairet, B. Rotenberg, and M. Salanne, Chemical Reviews **122**, 10860 (2022).

[8] J. I. Siepmann and M. Sprik, The Journal of Chemical Physics **102**, 511 (1998).

[9] L. Zeng, T. Wu, T. Ye, T. Mo, R. Qiao, and G. Feng, Nature Computational Science **1**, 725 (2021).

[10] C. Merlet, B. Rotenberg, P. A. Madden, P.-L. Taberna, P. Simon, Y. Gogotsi, and M. Salanne, Nature Materials **11**, 306 (2012).

[11] S. Bi *et al.*, Nature Materials **19**, 552 (2020).

[12] T. Dufils, G. Jeanmairet, B. Rotenberg, M. Sprik, and M. Salanne, Physical Review Letters **123**, 195501 (2019).

[13] S.-J. Shin, D. H. Kim, G. Bae, S. Ringe, H. Choi, H.-K. Lim, C. H. Choi, and H. Kim, Nature Communications **13**, 174 (2022).

[14] L. Scalfi, T. Dufils, K. G. Reeves, B. Rotenberg, and M. Salanne, The Journal of Chemical Physics **153**, 174704 (2020).

[15] R. Sundararaman, D. Vigil-Fowler, and K. Schwarz, Chemical Reviews **122**, 10651 (2022).

[16] A. Serva, L. Scalfi, B. Rotenberg, and M. Salanne, The Journal of Chemical Physics **155**, 044703 (2021).

[17] J.-B. Le, Q.-Y. Fan, J.-Q. Li, and J. Cheng, Science Advances **6**, eabb1219 (2020).

[18] N. D. Lang and W. Kohn, Physical Review B **7**, 3541 (1973).

[19] A. J. Stone and M. Alderton, Molecular Physics **56**, 1047 (1985).

[20] M. Brack, Reviews of Modern Physics **65**, 677 (1993).

[21] T. Pajkossy and D. Kolb, Electrochemistry Communications **5**, 283 (2003).

[22] N. Mott and R. Watts-Tobin, Electrochimica Acta **4**, 79 (1961).
11